\documentstyle[aps,tighten]{revtex}
\begin{document}
\preprint{EFEI-FIS}
\draft
\newcommand{\Fstar}{\raisebox{.2ex}{$\stackrel{*}{F}$}{}}
\newcommand{\astar}{\raisebox{-0.1ex}{$\stackrel{*}{a}$}{}}
\renewcommand{\thefootnote}{\fnsymbol{footnote}}
\title{Light propagation in non-linear electrodynamics}
\author{%
V. A. De Lorenci$^{1}$%
\protect\thanks{Electronic mail: \tt lorenci@cpd.efei.br}, 
R. Klippert$^{2,3}$%
\protect\thanks{Electronic mail: \tt klippert@icra.it}, 
M. Novello$^{2}$%
\protect\thanks{Electronic mail: \tt novello@lafex.cbpf.br} 
and J. M. Salim$^{2}$%
\protect\thanks{Electronic mail: \tt jsalim@lafex.cbpf.br}%
} 
\address{$^{1}$
Instituto de Ci\^encias - Escola Federal de Engenharia de Itajub\'a \\
Av. BPS 1303 Pinheirinho, 37500-000 Itajub\'a, MG -- Brazil} 
\address{$^{2}$
Centro Brasileiro de Pesquisas F\'{\i}sicas\\
Rua Dr.\ Xavier Sigaud 150 Urca, 
22290-180 Rio de Janeiro, RJ -- Brazil} 
\address{$^{3}$%
International Center for Relativistic Astrophysics \\
Piazza della Repubblica 10, 65100 Pescara -- Italy}

\date{\today}
\maketitle

\begin{abstract}
\hfill{\small\bf Abstract}\hfill\smallskip
\par
Working on the approximation of low frequency, we present
the light cone conditions for a class of theories constructed
with the two gauge invariants of the Maxwell field without 
making use of average over polarization states. Different
polarization states are thus identified describing 
birefringence phenomena. We make an application of the
formalism to the case of Euler-Heisenberg effective 
Lagrangian and well know results are obtained. 
\end{abstract}
\renewcommand{\thefootnote}{\arabic{footnote}}

\section{Introduction}

It is now a well established result that the velocity of the 
electromagnetic waves has its value dependent on the vacuum
polarization states.  Indeed, such polarization effects appear
when a strong field (electrodynamical critical field: 
$E_{\scriptscriptstyle cr}
= B_{\scriptscriptstyle cr} = m^2c^2/e\hbar \approx 1.3\times
10^{18}V/m \approx 4.4\times 10^{13}G$) is produced in some 
region of space. The most important
consequence of this fact consists in the birefringence effect:
the velocity of wave propagation depending on the wave polarization.
A experimental method to detect the vacuum birefringence induced
by a magnetic field was proposed in $1979$ by E. Iacoppini and E.
Zavattini \cite{Iacoppini}. In the experimental context, it is
worth to mention the work of D. Bakalov {\it et al} \cite{Bakalov}, 
where optical techniques is used to detect birefringence in the
presence of a strong magnetic field (PVLAS experiment). 
The theoretical description of nonlinear effects on light
propagation was studied long before by Z. Bialynicka-Birula and
I. Bialynicki-Birula \cite{Birula}, where was calculated the probability of
the photon splitting in an external electromagnetic field. The
same problem was extensively studied by S.L. Adler \cite{Adler}.
Other beautiful new results on vacuum polarization phenomena in
nontrivial vacua, including curved spacetimes, can be found in the
works of J.I. Latorre, P. Pascual and R. Tarrach \cite{Latorre},
I.T. Drumond and S.J. Hathrell \cite{Drumond}, G.M. Shore 
\cite{Shore} and others. See for instance the works of
K. Scharnhorst \cite{Scharnhorst} and G. Barton \cite{Barton}, 
where the problem of photon propagation between parallel mirrors
is worked out.

Recently, W. Dittrich and H. Gies \cite{Dittrich}, within the
approach of the geometrical optics, derived the light cone
conditions for a class of homogeneous nontrivial QED vacua
using the rule of average over polarization states. They generalized
some results previously obtained by Latorre and others, in
particular the so called ``unified formula'' \cite{Latorre}. Indeed,
such unified formula was identified by Latorre {\it et al} for
several modified QED vacua and proved by Dittrich and Gies for
certain cases. The use of the above mentioned average procedure
excludes from their formalism the possibility to analyse the
important phenomena of birefringence.

In this paper we deal with a class of Lagrangians
depending on the two Lorentz and gauge invariants of the Maxwell
field 
\begin{equation}
L = L(F^{\mu\nu}F_{\mu\nu},F^{\mu\nu}\Fstar_{\mu\nu})
\end{equation}
and working on the approximation of soft photons (the wavelength
of propagating wave is large compared to the Compton wavelength), 
we present the light cone conditions for local theories
of gauge invariant spin-one fields  
without making use of average over polarization states.  
Recent results by Dittrich and Gies 
are thus generalized in our approach, whose main contribution relies 
on the study of birefringence effects in a unified formalism.
The polarization problem is worked out and the dispersion law is
obtained, showing that there are a different polarization mode
associated to each velocity of wave propagation. Finally, we apply
the formalism to Euler-Heisenberg Lagrangian and some known results
are derived in the context of the present formalism. We set
the units $c=1=\hbar$.

\section{Non-linear spin-one theories}

Instead of calculating light cone conditions for each particular theory, 
we make use here of a general formalism, applicable to any Lagrangian based 
local theory describing gauge invariant spin-one fields that
can be constructed with the two invariants of the Maxwell field.  
We denote the electromagnetic field strength by the anti-symmetric 
2-rank tensor $F_{\mu\nu}$, and its dual is defined as
\begin{equation}
\label{6}
\Fstar_{\alpha\beta} = \frac{1}{2}\eta_{\alpha\beta}{}^{\sigma\tau}F_{\sigma\tau}.
\end{equation}
Let us set the only two local and gauge invariant scalar fields 
$F$ and $G$ associated with $F_{\mu\nu}$ by
\begin{mathletters}
\begin{eqnarray}
\label{1}
F&=&F^{\mu\nu}F_{\mu\nu}\\
\label{2}
G&=&F^{\mu\nu}\Fstar_{\mu\nu}.
\end{eqnarray}
\end{mathletters}
In order to achieve more
simplicity, we work in Minkowski spacetime employing a Cartesian coordinate
system. Thus, the background metric will be represented
by $\eta_{\mu\nu}$, which is defined by $\rm{diag}(+1,-1,-1,-1)$.
We defined the completely anti-symmetric tensor 
$\eta^{\alpha\beta\mu\nu}$
($\eta^{0123} = 1$), and set the notation 
$L_{X}=\partial L/\partial X$,
where the variable $X$ stands for any 
monomial on the field invariants. 
The gauge invariant density of Lagrangian of electrodynamics 
is an arbitrary function of $F$ and $G$:
\begin{equation}
\label{8}
L = L(F,G).
\end{equation}
From the minimal action principle we get the equation of motion
\begin{equation}
\label{9}
\left(L_F F^{\mu\nu} + L_G \Fstar^{\mu\nu}\right){}_{,\nu} = 0
\end{equation}
where a comma denotes partial derivatives 
with respect to the Cartesian coordinates.  
Using relations
\begin{math}
F_{,\nu} = 2F^{\alpha\beta}F_{\alpha\beta,\nu} 
\end{math} and 
\begin{math}
G_{,\nu} = 2F^{\alpha\beta}\Fstar_{\alpha\beta,\nu} 
\end{math}
in equation (\ref{9}) we obtain:
\begin{equation}
\label{20}
2N^{\mu\nu\alpha\beta}F_{\alpha\beta ,\nu} 
+ L_F F^{\mu\nu}{}_{,\nu} = 0
\end{equation} 
where we introduced the 4-rank tensor
$N^{\mu\nu\alpha\beta}$ through 
\begin{eqnarray}
N^{\mu\nu\alpha\beta} \doteq 
L_{FF}F^{\mu\nu}F^{\alpha\beta} 
+ L_{GG} \Fstar^{\mu\nu}\Fstar^{\alpha\beta}
+ L_{FG}\left( 
F^{\mu\nu}\Fstar^{\alpha\beta} +
\Fstar^{\mu\nu}F^{\alpha\beta}\right).
\label{21}
\end{eqnarray}
Additionally, the field strength $F_{\mu\nu}$ must satisfy the 
Bianchi identity.

Let us now turn our attention to the expressions which represent 
the light cone conditions for such general non-linear spin-one theory.

\section{Wave propagation}
\label{sec-light}

In this section we analyse the propagation of linear shock-waves 
associated with the discontinuities of the field 
in the limit of geometrical optics \cite{Hadamard}. 
Let us consider a surface of discontinuity $\Sigma$ defined by
\begin{equation}
z(x^{\mu}) = 0.
\end{equation}
Whenever $\Sigma$ is a global surface, it divides the spacetime 
in two distinct regions $U^-$ and $U^+$ ($z<0$ and $z>0$, respectively). 
Given an arbitrary function of the coordinates, $f(x^\mu)$, 
we define its discontinuity on $\Sigma$ as
\begin{equation}
\label{24}
\left[f(x^{\alpha})\right]_{\Sigma} \doteq \lim_{\{P^\pm\}\rightarrow P}
\left[f(P^+) - f(P^-)\right]
\end{equation}
where $P^+,\,P^-$ and $P$ belong to $U^+,\,U^-$ and $\Sigma$ respectively.
Applying the conditions \cite{Hadamard} of discontinuities 
for the tensor field $F_{\mu\nu}$ and its derivatives we set
\begin{mathletters}
\label{25}
\begin{eqnarray}
\label{25a}
\left[F_{\alpha\beta}\right]_{\Sigma} &=& 0 \\
\label{25b}
\left[F_{\alpha\beta ,\lambda}\right]_{\Sigma} &=& f_{\alpha\beta}k_{\lambda}
\end{eqnarray}
\end{mathletters}
where $f_{\alpha\beta}$ represents the discontinuities of the field on
the surface $\Sigma$ and $k_{\lambda}$ is the wave propagation 4-vector.
The discontinuity of the Bianchi identity yields
\begin{equation}
\label{26}
f_{\alpha\beta} k_{\lambda} + f_{\beta\lambda}k_{\alpha} +
f_{\lambda\alpha} k_{\beta} = 0.  
\end{equation} 
In order to obtain scalar relations, we consider 
the product of equation (\ref{26}) with
$F^{\alpha\beta}k^{\lambda}$  and with $\Fstar^{\alpha\beta}k^{\lambda}$,
which leads to \begin{mathletters}
\label{k2z1}
\begin{eqnarray}
\label{31}
\zeta k^{\lambda}k_{\lambda} &=& -2
F^{\alpha\beta}f_{\beta\lambda}k^{\lambda}k_{\alpha}\\
\label{32}
\zeta^* k^{\lambda}k_{\lambda} &=& -2
\Fstar^{\alpha\beta}f_{\beta\lambda}k^{\lambda}k_{\alpha}
\end{eqnarray}
\end{mathletters}
where we have introduced the scalar quantities
$\zeta$ and $\zeta^*$ as
\begin{mathletters}
\label{zeta}
\begin{eqnarray}
\label{29}
\zeta  &\doteq& F^{\alpha\beta}f_{\alpha\beta} \\
\label{30}
\zeta^* &\doteq& \Fstar^{\alpha\beta}f_{\alpha\beta}.
\end{eqnarray}
\end{mathletters}
In the same way, 
taking the discontinuities of the field equation (\ref{20}), we get 
\begin{equation}
\label{34}
f_{\beta\lambda} k^{\lambda} =
-\frac{2}{L_F}N_{\beta}{}^{\mu\nu\rho}f_{\nu\rho}k_{\mu}.
\end{equation}
Introducing relation (\ref{34}) in equations (\ref{k2z1}) 
and making use of the useful identities 
between the tensor field $F_{\mu\nu}$ and its dual:
\begin{mathletters}
\label{side}
\begin{eqnarray}
\Fstar_{\mu\alpha}F^{\alpha}{}_{\nu} &\equiv&
-\frac{1}{4}G\eta_{\mu\nu} \label{39} \\
F_{\mu\alpha}F^{\nu\alpha} - \Fstar_{\mu\alpha}
\Fstar^{\nu\alpha}
&\equiv& \frac{1}{2}F\delta_{\mu}{}^{\nu}
\label{41} 
\end{eqnarray}
\end{mathletters}
we obtain
\begin{mathletters}
\label{zk2}
\begin{eqnarray}
\zeta k^2 &=& \frac{4}{L_{F}}F^{\mu\nu}F^{\tau}{}_{\mu}k_{\nu}k_{\tau}
\left(L_{FF}\zeta + L_{FG}\zeta^*\right)
- \frac{G}{L_F}k^2\left(L_{FG}\zeta + L_{GG}\zeta^*\right)
\label{43}\\
\zeta^* k^2 &=& \frac{4}{L_{F}}F^{\mu\nu}F^{\tau}{}_{\mu}k_{\nu}k_{\tau}
\left(L_{FG}\zeta + L_{GG}\zeta^*\right)
- \frac{G}{L_F}k^2\left(L_{FF}\zeta + L_{FG}\zeta^*\right)
+\frac{2F}{L_F}k^2\left(L_{FG}\zeta + L_{GG}\zeta^*\right)
\label{44}
\end{eqnarray}
\end{mathletters}
where $k^2 \doteq \eta^{\mu\nu}k_{\mu}k_{\nu}$.

In order to find the equation that represents the propagation 
of the field discontinuities, we seek for a master relation 
which should be independent of the quantities $f_{\alpha\beta}$ 
(that is, independent of $\zeta$ and $\zeta^*$).
There is a simple way to  achieve such goal. 
We firstly isolate the common term 
\begin{math}
F^{\mu\nu}F^{\tau}{}_{\mu}k_{\nu}k_{\tau}
\end{math}
which appears in both equations (\ref{43}) and (\ref{44}). 
The difference of these equations can thus be written as  
\begin{eqnarray}
\frac{\zeta k^2}{L_{FF}\zeta + L_{FG} \zeta^*} -
\frac{\zeta^* k^2}{L_{FG}\zeta + L_{GG} \zeta^*} =
-\frac{2F k^2}{L_F}
+ \frac{G k^2}{L_{F}}\left(
\frac{L_{FF}\zeta + L_{FG} \zeta^*}{L_{FG}\zeta + L_{GG} \zeta^*} -
\frac{L_{FG}\zeta + L_{GG} \zeta^*}{L_{FF}\zeta + L_{FG} \zeta^*} 
 \right).
\label{47}
\end{eqnarray}
Assuming%
\footnote{Note that $k^2=0$ represents the standard propagation 
which occurs for the linear theory $L=-F/4$.  We are interested here 
in the study of the possible deviations from this simple case.} 
$k^2 \neq 0$, we obtain an algebraic linear relation 
\begin{equation}
\Omega_1 {\zeta^*}^2 + \Omega_2 \zeta \zeta^* +\Omega_3 \zeta^2=0
\label{49}
\end{equation}
between $\zeta$ and $\zeta^*$, where we defined
\begin{mathletters}
\label{Omegas}
\begin{eqnarray}
\Omega_1 &\doteq& -L_{FG} +2\frac{F}{L_F}L_{FG}L_{GG} - \frac{G}{L_F}
L_{FG}^2 + \frac{G}{L_F}L_{GG}^2 \label{50} \\
\Omega_2 &\doteq& L_{GG}-L_{FF} +2\frac{F}{L_F}L_{FF}L_{GG} + 2\frac{F}{L_F}
L_{FG}^2
- 2\frac{G}{L_F}L_{FF}L_{FG} + 2\frac{G}{L_F}L_{FG}L_{GG} 
\label{51} \\
\Omega_3 &\doteq& L_{FG} +2\frac{F}{L_F}L_{FF}L_{FG} - \frac{G}{L_F}
L_{FF}^2 + \frac{G}{L_F}L_{FG}^2. 
\label{52} 
\end{eqnarray} 
\end{mathletters}
Solving the quadratic equation (\ref{49}) for $\zeta^*$ we obtain
\begin{equation}
\zeta^* = \Omega_{{\scriptscriptstyle \pm}} \zeta
\label{54}
\end{equation}
with
\begin{equation}
\Omega_{{\scriptscriptstyle \pm}} = \frac{-\Omega_2 \pm \sqrt{\Omega_2^2 - 
4\Omega_1\Omega_3}}{2\Omega_1}.
\label{55}
\end{equation}
Using this solution into equation (\ref{43}) 
and assuming $\zeta \neq 0$ it results the 
following light cone conditions for spin-one fields:
\begin{eqnarray}
&&\left[ 1 + \frac{G}{L_F}\left(L_{FG}+
\Omega_{{\scriptscriptstyle \pm}}   L_{GG}\right)
\right]k^2
-\frac{4}{L_{F}}
\left(L_{FF}+ \Omega_{{\scriptscriptstyle \pm}}  
L_{FG}\right)F^{\mu\nu}F^{\tau}{}_{\mu}k_{\nu}k_{\tau} = 0.
\label{57a}
\end{eqnarray}
In a similar way for equation (\ref{44}) we obtain 
\begin{eqnarray}
&&\left[ \Omega{\scriptscriptstyle \pm} 
- \frac{2F}{L_F}\left(L_{FG} + \Omega_{\scriptscriptstyle \pm}L_{GG}\right)
\right. + \left.\frac{G}{L_F}\left(L_{FF} + \Omega_{\scriptscriptstyle
\pm}  L_{FG}\right) \right]k^2 
-\frac{4}{L_{F}}\left(L_{FG}+ \Omega_{{\scriptscriptstyle \pm}}  
L_{GG}\right)F^{\mu\nu}F^{\tau}{}_{\mu}k_{\nu}k_{\tau} = 0.
\label{57b}
\end{eqnarray}
After some algebra, 
it can be shown that equation (\ref{57b}) is identical to (\ref{57a}).  

The light cone conditions following from such procedure 
lead to two possible paths of propagation, according with
the double solutions $\Omega_{\scriptscriptstyle \pm}$. 
These two conditions are related to distinct polarizations modes,
as we will see in the next section, 
indicating the possibility of birefringence.  
This effect depends upon the particular theory we shall consider.

In a more appealing form, we can present expression (\ref{57a}) as
\begin{eqnarray}
k^2=4\frac{L_{FF}+ \Omega_{{\scriptscriptstyle \pm}}  
L_{FG}}{L_{F} + G\left(L_{FG}+
\Omega_{{\scriptscriptstyle \pm}}L_{GG}\right)}
F^{\lambda\mu}F^{\nu}{}_{\lambda}k_{\mu}k_{\nu}.
\label{57}
\end{eqnarray}
Since $k^2 = \omega_{o}^2 - |\vec{k}|^2$, the phase velocity
$\omega_{o}/|\vec{k}|\equiv v$ of the propagating light is
found to be
\begin{eqnarray}
v^2_{\scriptscriptstyle \pm}=1-4\frac{L_{FF}+ 
\Omega_{{\scriptscriptstyle
\pm}}   L_{FG}}{L_{F} + G\left(L_{FG}+
\Omega_{{\scriptscriptstyle \pm}}L_{GG}\right)}
F^{\lambda\mu}F^{\nu}{}_{\lambda}k_{\mu}k_{\nu}.
\label{phasepn}
\end{eqnarray}
This equation indicates that the familiar case $k^2=0$, 
which occurs for linear electrodynamics $L=-F/4$, 
is also possible for more complicated situations, 
as for those theories satisfying the relation
$L_{FF} + \Omega_{{\scriptscriptstyle \pm}}L_{FG}=0$.  
Solutions of this equation brings up the form 
of particular Lagrangians for which $k^2=0$, 
despite it was previously assumed that $k^2\neq 0$.  
A simple solution of this equation can be obtained by setting 
$L = -F/4 + f(G)$, where $f(G)$ is an arbitrary function 
of the invariant $G$. Another example where $k^2=0$ occurs
consists in the nonlinear Lagrangian of the N. Born and 
L. Infeld \cite{Born}. 
We are not interested here  on the analysis of such theories.

The light cone conditions (\ref{57}) can also be expressed 
in terms of the energy-momentum tensor of the non-linear field 
\begin{equation}
T_{\mu\nu} = -4 L_{F}\, F_{\mu}{}^{\alpha} \, F_{\alpha\nu} - 
\left( L - G \, L_{G} \right) \,\eta_{\mu\nu}
\label{Tmunu}
\end{equation}
as
\begin{equation}
k^2  = - Q_{\scriptscriptstyle \pm} T^{\mu\nu}k_{\mu}k_{\nu}
\label{kAB}
\end{equation}
where we defined the quantities 
\begin{eqnarray}
Q_{\scriptscriptstyle \pm} \doteq \frac{\left( L_{FF} + 
\Omega_{\scriptscriptstyle \pm} L_{FG} \right)}
{{L_F}^2 + G L_F
\left( L_{FG} + \Omega_{\scriptscriptstyle \pm} L_{GG} \right)
+ \left( L_{FF} + \Omega_{\scriptscriptstyle \pm} 
L_{FG} \right) \left( L - G L_{G} \right)}.
\label{Q}
\end{eqnarray}
Thus, in terms of the energy momentum tensor, 
the phase velocity for each
propagation mode,  corresponding to the solutions of
$\Omega_{\scriptscriptstyle \pm}$, are given by 
\begin{equation}
v_{\pm}^2 = 1 - Q_{\scriptscriptstyle \pm} T^{\mu\nu}n_{\mu}n_{\nu}
\label{vgeral}
\end{equation}
where we introduced the quantity 
$n_{\mu} \doteq k_{\mu}/|\vec{k}|$
specifying the direction of wave propagation.  

In the literature one usually makes use of an average over polarization 
states, which can directly be obtained from our formalism in the
form $v = (v_+ + v_-)/2$.
This represents the velocity of propagation for the average mode.  

The results we had obtained in this section are applied 
for spin-one theories which is set by the Lagrangian function 
defined by (\ref{8}), in the approximation of soft photons.  
The light cone
conditions,  here presented in terms of the field strength by (\ref{57}) 
or else in terms of the energy-momentum tensor by (\ref{kAB}), 
are useful in a variety of situations, 
and particularly for the study of birefringence phenomena.
Let us now analyse in more details the problem of polarization.

\section{Polarization}

The most general decomposition for a skew-symmetric tensor is 
$f_{\alpha\beta}=(A_\alpha B_\beta-A_\beta B_\alpha)
+(C_\alpha D_\beta-C_\beta D_\alpha)$, where the vectors 
$A_\alpha$, $B_\alpha$, $C_\alpha$, $D_\alpha$ are arbitrary.  
For the case where $f_{\alpha\beta}$ is the wave propagation tensor 
given by equation (\ref{25b}), for which equation (\ref{26}) applies,
it follows that the above decomposition simplifies to 
\begin{equation}
\label{fepsilon}
f_{\alpha\beta} = a (\epsilon_\alpha k_\beta-\epsilon_\beta k_\alpha)
\end{equation}
where $a$ is the strength of the wavelet and $\epsilon_\beta$
represents the polarization vector.  
 
The field equations impose restrictions on the possible states
determined by such vector.
Introducing (\ref{fepsilon}) in (\ref{34}) and assuming $a\neq 0$ we have 
\begin{equation}
\label{k2e}
k^2\,\epsilon^\mu = -\frac{4}{L_F}
(N^{\mu\alpha}{}_{\nu\beta}k_\alpha k^\beta)\,\epsilon^\nu.
\end{equation}
 The $k^2=0$ case includes the linear propagation regime,
where the polarization modes are well known.
In this case equation (\ref{k2e}) states that 
$N^{\mu\alpha}{}_{\nu\beta}k_\alpha k^\beta \epsilon^\nu = 0$.
The $k^2\neq 0$ case can be treated defining 
a symmetric tensor $Z_{\mu\nu}$ by   
\begin{equation}
\label{Z}
Z^\mu{}_\nu \doteq \delta^\mu{}_\nu 
+\frac{4}{L_F k^2}N^{\mu\alpha}{}_{\nu\beta}k_\alpha k^\beta.
\end{equation}
With it we can write (\ref{k2e}) as an eigenvector equation
\begin{equation}
\label{Ze}
Z^\mu{}_\nu \epsilon^\nu = 0.
\end{equation}
The solutions of equation (\ref{Ze}) (eigenvectors of $Z^\mu{}_\nu$) 
represent the dynamically allowed polarization modes.
The definition of  $N^{\mu\nu\alpha\beta}$, from equation (\ref{21}),
leads us to conclude that the tensor structure of 
$Z^{\mu}{}_{\nu}$ can be completely determined by the electromagnetic
tensor, its dual and the wave vector $k^{\mu}$.
Hence, the general solution for the eigenvector problem can be
achieved by expanding $\epsilon_{\mu}$ 
as a linear combination of the following linearly
independent vectors: 
\begin{equation}
F^{\mu\nu}k_{\nu}\equiv a^{\mu},\,\,\,
\Fstar^{\mu\nu}k_{\nu}\equiv \astar^{\mu}, \,\,\, 
F^{\mu\alpha}F_{\alpha\nu}k^{\nu}\equiv b^{\mu}, \,\,\, 
k^{\mu}.
\end{equation} 
Thus, the polarization vector takes the form
\begin{equation}
\epsilon_{\mu} = \alpha a^{\mu} + \beta \astar^{\mu}
+ \gamma k^{\mu} + \delta b^{\mu}.
\label{epsilon}
\end{equation}
Introducing (\ref{epsilon}) in (\ref{Ze}) and taking the
products, results
\begin{eqnarray}
&&\left\{\alpha\left[ \frac{L_F}{4}k^2 + L_{FF}a^2 + L_{FG}a^{\mu}
\astar_{\mu}\right] + \beta\left[L_{FF}a^{\mu}
\astar_{\mu} + L_{FG}\astar^{\mu}
\astar_{\mu}\right]\right\}a^{\nu}
\nonumber \\
&+&\left\{\alpha\left[L_{FG}a^2 + L_{GG}a^{\mu}
\astar_{\mu}\right] + \beta\left[\frac{L_F}{4}k^2 +
L_{FG}a^{\mu}\astar_{\mu} + L_{GG}\astar^{\mu}
\astar_{\mu}\right]\right\}\astar^{\nu}
\nonumber\\
&+& \gamma [0]k^{\nu} + \delta \left[\frac{L_F}{4}k^2\right]b^{\nu}=0.
\label{e12}
\end{eqnarray}
In order to satisfy the above equation, the coefficients of each
independent vector must be null, resulting in
\begin{mathletters}
\label{f1-4}
\begin{eqnarray}
\alpha\left[ \frac{L_F}{4}k^2 + L_{FF}a^2 + L_{FG}a^{\mu}
\astar_{\mu}\right] + \beta\left[L_{FF}a^{\mu}
\astar_{\mu} + L_{FG}\astar^{\mu}
\astar_{\mu}\right] &=& 0
\label{f1} \\
\alpha\left[L_{FG}a^2 + L_{GG}a^{\mu}
\astar_{\mu}\right] + \beta\left[\frac{L_F}{4}k^2 +
L_{FG}a^{\mu}\astar_{\mu} + L_{GG}\astar^{\mu}
\astar_{\mu}\right] &= &0
\label{f2}\\
\gamma &=& {\rm arbitrary} 
\label{f3}\\
\delta &=& 0.
\label{f4}
\end{eqnarray}
\end{mathletters}
Therefore, from (\ref{fepsilon}), $\gamma k_{\mu}$ does not
contribute to $f_{\alpha\beta}$, and we shall not consider it
in any further. Using the relations (\ref{side}) yields
\begin{mathletters}
\label{f5-6}
\begin{eqnarray}
a^{\mu}\astar_{\mu} & =& \frac{1}{4}Gk^2
\label{f5} \\
\astar^{\mu}\astar_{\mu} & = & 
a^2 - \frac{1}{2}F k^2.
\label{f6}
\end{eqnarray}
\end{mathletters}
Substituting these results in equations (\ref{f1-4}) and solving
the system for $k^2$ we obtain the dispersion laws 
\cite{Birula,Boillat} for both polarization modes, which are
described by the vectors 
$\epsilon^{\mu}_{\scriptscriptstyle \pm}$. Indeed, it can be
shown that such dispersion relations are recovered from the light cone
conditions (\ref{57}) for all cases
known in the literature, ensuring
our previous statement concerning the relation between 
$\Omega_{\scriptscriptstyle \pm}$ with the two polarization 
states .

\section{Application to Euler-Heisenberg Lagrangian}

In this section we apply the previous results 
to retrieve, from our formalism, a particular case of birefringence 
presented in the literature \cite{Bakalov}.  
The most familiar non-linear case of electrodynamics is given by
Euler-Heisenberg \cite{Euler} effective action of QED\footnote{
The limit of weak-field from one loop approximated QED action.},
\begin{equation}
S = \int dx \left[-\frac{1}{4}F + \frac{\mu}{4}\left(F^2 +
\frac{7}{4}G^2\right)\right]
\label{euler}
\end{equation}
where $dx$ stands for the volume measure of the spacetime, 
and the quantum parameter $\mu$ is defined by
\begin{equation}
\mu \doteq \frac{2\alpha^2}{45 m_{\scriptscriptstyle e}^4}.
\label{mu}
\end{equation}
The light cone condition for each propagation mode can be directly 
obtained from expression (\ref{57}) as 
m\begin{mathletters}
\label{k+-}
\begin{eqnarray}
k^2 &=& -8\mu F^{\alpha\mu}F_{\mu}{}^{\beta}k_{\alpha}k_{\beta}
\label{k+}\\
k^2 &=& -14\mu F^{\alpha\mu}F_{\mu}{}^{\beta}k_{\alpha}k_{\beta}.
\label{k-}
\end{eqnarray}
\end{mathletters}
From equation (\ref{Q}), in the required order of approximation
we obtain
\begin{mathletters}
\label{Q+-}
\begin{eqnarray}
Q_{\scriptscriptstyle +} &=& 8\mu
\label{Q+}\\
Q_{\scriptscriptstyle -} &=& 14\mu.
\label{Q-}
\end{eqnarray}
\end{mathletters}
Thus, in terms of the energy-momentum tensor,
\begin{mathletters}
\label{kT}
\begin{eqnarray}
k^2 &=& -8\mu T^{\alpha\beta}k_{\alpha}k_{\beta}
\label{k+T}\\
k^2 &=& -14\mu T^{\alpha\beta}k_{\alpha}k_{\beta}.
\label{k-T}
\end{eqnarray}
\end{mathletters}  
The phase velocities (\ref{phasepn}) corresponding to 
(\ref{k+-}) reduce to 
\begin{mathletters}
\label{v+-}
\begin{eqnarray}
v_{+} &=& 1 - 4\mu F^{\alpha\mu}F_{\mu}{}^{\beta}n_{\alpha}n_{\beta}
\label{v+}\\
v_{-} &=& 1 - 7\mu F^{\alpha\mu}F_{\mu}{}^{\beta}n_{\alpha}n_{\beta}.
\label{v-}
\end{eqnarray}
\end{mathletters}
These velocities correspond to two polarization states, showing
that birefringence
effect occurs in Euler-Heisenberg model, as it is well known in 
the literature.
From the polarization sum rule we obtain
\begin{equation}
v^2 = 1 - 11\mu F^{\alpha\mu}F^{\beta}{}_{\mu}n_{\alpha}n_{\beta}.
\label{sum}
\end{equation}
This result was presented in the paper of  G. M. Shore \cite{Shore},
and more recently by W. Dittrich and H. Gies \cite{Dittrich}.

\section{Conclusion}

In this work we derived light cone conditions for 
a class of local and gauge invariant spin-one 
field theory constructed with the two invariants of the Maxwell
field in the approximation of low frequency 
without making use of the average over polarization modes.
In our formalism we took the different polarization states into
account explicitly and one achieved a
generalization of the wave propagation formula, which can be
applied to the analysis  of birefringence effects in a unified way. 
In order to illustrate its
applications, we exhibited how to  obtain some results of
quantum electrodynamics  concerning the wave velocity dependence on
polarization states. 

An interesting continuation of this work would be the analysis of
the wave propagation equations as an effective modification of Minkowskian
geometry for each polarization direction, due to 
non-linear effects. Investigations on this topic had 
already been performed (see reference \cite{Plebanski}, and
more recently \cite{nos}). 
The use of the formalism presented here to the case of wave 
propagation in non-linear material media, also deserves further 
investigations.

\acknowledgements
This work was supported by {\em Conselho Nacional
de Desenvolvimento Cient\'{\i}fico e Tecnol\'ogico} (CNPq)
of Brazil.


\begin{references}

\bibitem{Iacoppini}
E. Iacopini and E. Zavattini, 
{\em Phys.\ Lett.} {\bf B 85}, 151 (1979).

\bibitem{Bakalov}
D. Bakalov, 
{\em Nucl.\ Phys.} {\bf B 35} (Proc.\ suppl.), 180 (1994).

\bibitem{Birula}Z. Bialynicka-Birula and I. Bialynicki-Birula, 
{\em Phys.\ Rev.} {\bf D 2}, 2341 (1970).

\bibitem{Adler}
S. L. Adler, {\em Ann. Phys.} {\bf 67}, 599 (1971).

\bibitem{Latorre}
J. I. Latorre, P. Pascual and R. Tarrach, 
{\em Nuclear Phys.} {\bf B 437}, 60 (1995).

\bibitem{Drumond} 
I. T. Drummond and S. J. Hathrell,
{\em Phys.\ Rev.} {\bf D 22}, 343 (1980).

\bibitem{Shore} 
G. M. Shore, 
{\em Nuclear Phys.} {\bf B 460}, 379 (1996);
R. D. Daniels and G. M. Shore, 
{\em Nuclear Phys.} {\bf B 425}, 634 (1994).

\bibitem{Scharnhorst} 
K. Scharnhorst, 
{\em Phys. Lett.} {\bf B 236}, 354 (1990).

\bibitem{Barton} 
G. Barton, 
{\em Phys. Lett.} {\bf B 237}, 559 (1990).

\bibitem{Dittrich}
W. Dittrich and H. Gies, 
{\em Phys.\ Rev.} {\bf D 58}, 025004 (1998); 
{\it ibid,} {\em Phys.\ Lett.} {\bf B 431}, 420 (1998).

\bibitem{Hadamard} 
Y. Choquet-Bruhat, C. De Witt-Morette, M. Dillard-Bleick,
in {\it Analysis, Manifolds and Physics}, p.\ 455 
(North-Holland Publishing, N.Y., 1977); 
see also J. Hadamard, in {\it Le\c{c}ons sur la propagation 
des ondes et les \'equations de l'hydrodynamique}, 
(Ed. Hermann, Paris, 1903).

\bibitem{Born} M. Born, Nature, {\bf 132}, 282 (1933); 
Proc. Roy. Soc. {\bf A 143}, 410 (1934); 
M. Born and L. Infeld, Nature {\bf 133}, 63 (1934).

\bibitem{Euler}
W. Heisenberg and H. Euler, Z. Phys.\ {\bf 98}, 714 (1936); 
J. Schwinger, {\em Phys.\ Rev.} {\bf 82}, 664 (1951).

\bibitem{Boillat}
G. Boillat, {\em J. Math. Phys.} {\bf 11}, 941 (1970).

\bibitem{Plebanski}
J. Plebanski, in {\em Lectures on non-linear electrodynamics.} 
(Nordita, Copenhagen, 1968).

\bibitem{nos}
M. Novello, V. A. De Lorenci, J. M. Salim and R. Klippert,
{\em Phys. Rev.} {\bf D 61}, 45001 (2000).

\end{references}
\end{document}